\documentclass[10pt]{iopart}

\usepackage{graphicx}
\usepackage{dcolumn}
\usepackage{bm}

\usepackage{placeins}
\usepackage{graphicx}
\usepackage{color}
\usepackage{hyperref}
\usepackage{multirow}

\begin{document}
\title[Electrically tunable exchange splitting in bilayer graphene on Cr$_2$Ge$_2$Te$_6$]{Electrically tunable exchange 
splitting in bilayer graphene on monolayer \texorpdfstring{Cr$_2$X$_2$Te$_6$ with X=Ge, Si, and Sn.}{}}
\author{Klaus Zollner, Martin Gmitra and Jaroslav Fabian}
\address{Institute for Theoretical Physics, University of Regensburg, 93040 Regensburg, Germany\\
}
\ead{klaus.zollner@physik.uni-regensburg.de}
\begin{abstract}
We investigate the electronic band structure and the proximity exchange effect in bilayer graphene on a family
of ferromagnetic multilayers Cr$_2$X$_2$Te$_6$, X=Ge, Si, and Sn, with first principles methods. In each case
the intrinsic electric field of the heterostructure induces an orbital gap on the order of 10 meV in the graphene bilayer.
The proximity exchange is strongly band dependent. For example, in the case of Cr$_2$Ge$_2$Te$_6$, 
the low-energy valence band of bilayer graphene has exchange splitting of 8 meV, while the low energy conduction band's splitting
is 30 times less (0.3 meV). This striking discrepancy stems from the layer-dependent 
hybridization with the ferromagnetic substrate. Remarkably, applying a vertical electric field of a few V/nm reverses the exchange, 
allowing us to effectively turn ON and OFF proximity magnetism in bilayer graphene. Such a field-effect should be
generic for van der Waals bilayers on ferromagnetic insulators, opening new possibilities for spin-based devices.
\end{abstract}

\maketitle

\section{Introduction}
The properties of two-dimensional materials can be strongly altered by proximity effects
in van der Waals heterostructures. 
One prominent example is graphene, which experiences strong proximity spin-orbit coupling (SOC) effects 
when placed on transition metal dichalcogenides (TMDC) \cite{Gmitra2015:PRB, Gmitra2016:PRB}, 
and a giant field-effect SOC when bilayer graphene is used \cite{Gmitra2017:PRL}. 
Typical experimental structures like graphene/hBN/ferromagnet, showing very efficient 
spin injection \cite{Gurram2017:NC, Kamalakar2014:SR, Yamaguchi2013:APE, Kamalakar2016:SR, Fu2014:JAP}, also feature significant 
proximity exchange \cite{Zollner2016:PRB, Lazic2016:PRB}. 
These heterostructures are presently used in optospintronic devices \cite{Gmitra2015:PRB, Avsar2017:arXiv, Luo2017:NL} and also
for spin transport \cite{Kamalakar2014:SR, Fu2014:JAP, Gurram2017:NC, Kamalakar2016:SR, Yamaguchi2013:APE} 
in graphene spintronics \cite{Han2014:NN}.
However, when it comes to device realizations, bilayer graphene (BLG) offers many advantages, due
to the possibility of having an electrically tunable band gap \cite{Zhang2009:N, Oostinga2008:NM}, as well 
as the more precise control of the chemical potential \cite{Martin2008:NP,Rutter2011:NP}.

Graphene is a diamagnet. Can we make it ``ferromagnetic" 
so that it could be used for spintronics applications \cite{Zutic2004:RMP, Fabian2007:APS}? It turns out
that an effective way to achieve this is to place graphene on a ferromagnetic substrate (insulator or semiconductor,
to preserve the Dirac character of the bands and avoid substrate transport). Graphene then experiences 
proximity exchange coupling, typically in the range of $1-10$ meV \cite{Zollner2016:PRB, Lazic2016:PRB, Wang2015:PRL, Leutenantsmeyer2016:2dM, Mendes2015:PRL, Sakai2016:ACS, Swartz2012:ACS, Wei2016:NM, Dyrda2017:arXiv, Zhang2015:PRB, Song2017:arxiv, Haugen2008:PRB}, sometimes reaching even higher values~\cite{Zhang2015:SR, Yang2013:PRL, Hallal2017:2DM, Su2017:PRB}. 
Exchange proximity effects in graphene were observed by quantum anomalous Hall effect 
\cite{Wang2015:PRL}, magnetoresistance~\cite{Mendes2015:PRL} and nonlocal spin transport experiments~\cite{Leutenantsmeyer2016:2dM},
as has been confirmed theoretically \cite{Dyrda2017:arXiv, Zhang2015:PRB, Song2017:arxiv, Haugen2008:PRB, Zhang2015:SR, Su2017:PRB}.   

Several ferromagnetic insulators (FMI), like EuO and Yttrium-Iron-Garnet, 
have been considered in the context of graphene spintronics~\cite{Yang2013:PRL, Mendes2015:PRL, Su2017:PRB, Hallal2017:2DM, Singh2017:PRL}.
However transition-metal trichalcogenides Cr$_2$Si$_2$Te$_6$, Cr$_2$Ge$_2$Te$_6$ (CGT), and transition-metal trihalides CrI$_3$, and their monolayers are now extensively discussed in literature
as promising materials for low-dimensional spintronics because of their 
FMI ground state properties~\cite{Carteaux1991:JMM, Carteaux1995:JP, Williams2015:PRB, Casto2015:APL, Gong2017:Nat, Zhang2016:JJAP,
Siberchicot1996:JPC, Lin2016:JMCC, Chen2015:PRB, Liu2017:PRB, Sivadas2015:PRB, Li2014:JMC, Lin2017:PRB, Xing2017:2dM, Alegria2014:APL, Ji2013:JAP, Huang2017:Nat, Liu2016:PCCP, McGuire2017:Cryst}. 
These materials have attracted attention as substrates for 
topological insulators (TI)~\cite{Alegria2014:APL, Ji2013:JAP} and graphene~\cite{Zhang2015:PRB}, 
because ferromagnetic coupling is present. 
For example, CGT seems to be a nice platform for large scale epitaxial growth of TIs with high quality interfaces, 
where the interaction of TI surface states and ferromagnetism can be studied~\cite{Alegria2014:APL, Ji2013:JAP}. 
Moreover, first principles calculations show that monolayer graphene on CGT~\cite{Zhang2015:PRB} features
exchange splitting of about $10$~meV. It is then natural to try to use them for proximity effects in BLG.

For device operations, it is desirable to have electric control of the band structure and spin properties. We already know that BLG on 
a substrate can open a gap (similarly to having a vertical electric field applied). {\it Could we also control exchange
coupling in a similar way, simply applying an electric field?} One naturally expects that the proximity effect is strongest
in the BLG layer adjacent to the ferromagnetic substrate. Orbitals from this layer form, say, the valence band, which 
will exhibit strong exchange splitting. The conduction band, on the other hand, would have no or only weak splitting.
If an electric field is applied, the situation can be reversed, and now it would be the conduction band that 
has the largest exchange splitting. It has also been shown in model calculations, 
that BLG on a FMI can be a platform for field-effect magnetic 
or spin devices \cite{Michetti2010:NL, Michetti2011:PRB, Jatiyanon2016:CPB, Yu2011:PL, Rashidian2014:JP}. The combination of 
exchange and electric field leads to the control of the spin-dependent gap of the carriers and
perfect switching of the spin polarization is predicted \cite{Jatiyanon2016:CPB, Yu2011:PL}.
{\it How realistic is such an electrical control of proximity exchange splitting in BLG on real magnetic substrates? }
We answer this question by performing DFT calculations of BLG 
on monolayer Cr$_2$Si$_2$Te$_6$, CGT, and Cr$_2$Sn$_2$Te$_6$. Indeed, we show that 
the exchange splitting in a given band can be switched ON and OFF by a relatively modest vertical electric field. 
This opens venues for investigating magnetotransport in BLG, as well as for controlling the \textit{synthetic} magnetism
by gates. The analogy with spin-orbit valve effect \cite{Gmitra2017:PRL} is certainly there. But the exchange
valve is {\it fundamentally} different, in that it generates a magnetic moment in BLG, splits the two valleys equally,
and opens up a new platform for magnetotransport phenomena and spin filtering. There are now
recent experiments demonstrating an electric field control of ferromagnetism in 2D ferromagnets (such as CrI$_3$ \cite{Jiang2018:arxiv} and CGT \cite{Wang2018:arxiv}).  Controlling ``synthetic" magnetism
in BLG on such materials would be a great achievement.

In this paper we investigate, by performing first principles calculations, the proximity induced exchange
coupling in BLG on monolayer CGT, as well as on Cr$_2$Si$_2$Te$_6$ and Cr$_2$Sn$_2$Te$_6$ (the last has been theoretically
predicted~\cite{Zhuang2015:PRB} but not yet prepared experimentally). 
We find that the low energy bands of BLG show an indirect band gap of roughly
$17$~meV, which can be efficiently tuned by experimentally accessible electric fields of a few V/nm. 
The proximity induced exchange splitting of the valence band is giant, being around $8$~meV, $30$-times larger than
the exchange splitting in the conduction band. This large difference arises from the fact, that the valence band
is formed by non-dimer carbon atoms from the bottom graphene layer directly above CGT, where the proximity effect is strong,
while the conduction band is formed by non-dimer carbon atoms from the top graphene layer, where the proximity effect is weak
(in BLG with Bernal stacking, one pair of carbon atoms is vertically connected, which we call dimer, the other pair non-dimer).
The most interesting result is the switching of the exchange splitting from the valence to the conduction band, 
via an external electric field that counters the built-in field of the BLG/CGT heterostructure. 
\textit{A proximity exchange valve is realized.} We also include a brief discussion on possible device geometries 
based on the field-effect proximity exchange.

\section{Computational details and geometry}

\begin{figure}[htb]
\centering
	\includegraphics[width=0.9\columnwidth]{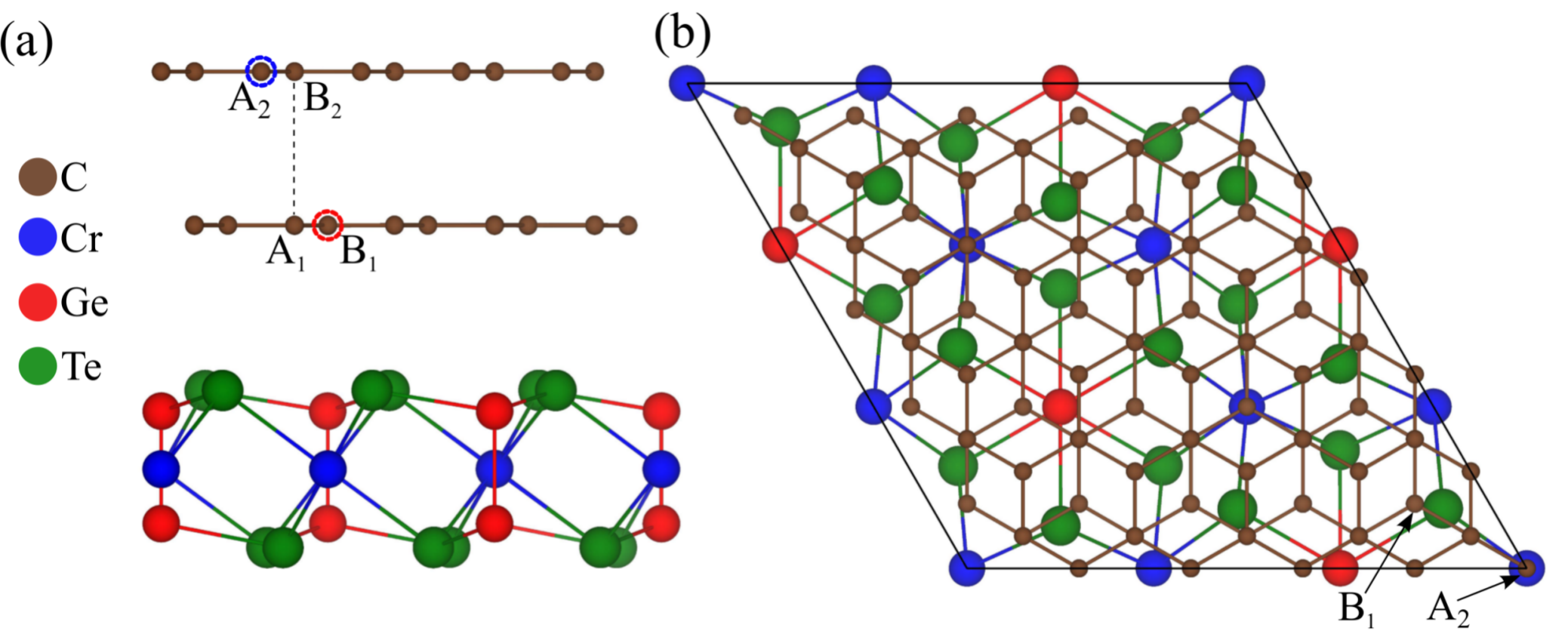}
	\caption{(Color online) (a) Side view of bilayer graphene on monolayer Cr$_2$Ge$_2$Te$_6$ (supercell)
		with labels for the different atoms. Labels A$_1$, B$_1$, A$_2$ and B$_2$ 
		indicate the sublattices of bilayer graphene in Bernal stacking. 
		Orbitals on non-dimer atoms B$_1$ and A$_2$
		form the low energy valence and conduction bands in the electronic structure 
		of bilayer graphene, with B$_1$ being closer to the substrate.
		(b) Top view of the atomic structure of
		bilayer graphene on monolayer Cr$_2$Ge$_2$Te$_6$, showing the positions of B$_1$ and A$_2$ above CGT.}
	\label{Fig:scheme}
\end{figure}

The heterostructure of BLG on CGT is shown in Fig.~\ref{Fig:scheme}, 
where a $5 \times 5$ supercell of BLG is placed on a $\sqrt{3} \times \sqrt{3}$ CGT supercell.
The considered heterostructure model contains 130 atoms in the unit cell. 
We keep the lattice constant of graphene unchanged at $a = 2.46$~\AA~and 
stretch the lattice constant of CGT by roughly 4\% from $6.8275$~\AA~\cite{Carteaux1995:JP} to $7.1014$~\AA.
Theoretical calculations predict that the tensile strain leaves the ferromagnetic ground state unchanged, but 
enhances the band gap and the Curie temperature of CGT~\cite{Li2014:JMC, Chen2015:PRB}.
In Fig.~\ref{Fig:scheme}(a) the side view of the structure shows the Bernal stacking of BLG, with an average
distance relaxed to $3.266$~\AA~between the graphene layers, in good agreement with experiment \cite{Baskin1955:PR}. 
The average distance between the lower graphene layer and CGT was relaxed to $3.516$~\AA, 
consistent with literature \cite{Zhang2015:PRB}. 

Previous calculations show that the relative alignment of 
graphene on CGT does not influence the electronic bands much, see Ref.~\cite{Zhang2015:PRB}, 
and we will only consider the supercell with stacking as in Fig.~\ref{Fig:scheme}. 
We will also not consider SOC in the calculation, since it was shown that proximity induced exchange in graphene,
caused by CGT, is one order of magnitude larger than proximity SOC~\cite{Zhang2015:PRB}.
The electronic structure calculations and structural relaxation of BLG on CGT were performed by 
means of density functional theory (DFT)~\cite{Hohenberg1964:PRB} within {Quantum ESPRESSO}~\cite{Giannozzi2009:JPCM}.
Self-consistent calculations were performed with the $k$-point sampling of $18\times 18\times 1$ to get converged results
for the proximity exchange splittings. 
We have performed open shell calculations that provide the 
spin polarized ground state with a collinear magnetization. 
Theory and experiment predict that CGT is a FMI with 
magnetic anisotropy favoring a magnetization perpendicular to 
the CGT-plane~\cite{Zhang2016:JJAP, Williams2015:PRB, Zhuang2015:PRB, 
Carteaux1991:JMM, Carteaux1995:JP, Casto2015:APL, Ji2013:JAP, Gong2017:Nat}. 
A Hubbard parameter of $U = 1$~eV was used for Cr $d$-orbitals, 
being in the range of proposed $U$ values especially for this compound \cite{Gong2017:Nat}.
The value results from comparison of DFT and experiment on the magnetic ground state of bulk CGT.
Other theoretical calculations report, that their results are qualitatively independent from the 
used $U$ values~\cite{Li2014:JMC, Sivadas2015:PRB}.
We used an energy cutoff for charge density of $500$~Ry, and
the kinetic energy cutoff for wavefunctions was $60$~Ry for the scalar relativistic pseudopotential 
with the projector augmented wave method \cite{Kresse1999:PRB} with the 
Perdew-Burke-Ernzerhof exchange correlation functional \cite{Perdew1996:PRL}.
For the relaxation of the heterostructures, we added 
van-der-Waals corrections \cite{Grimme2006:JCC,Barone2009:JCC} and used 
quasi-newton algorithm based on trust radius procedure. 
In order to simulate quasi-2D systems the vacuum of $20$~\AA~was used 
to avoid interactions between periodic images in our slab geometry.
Dipole corrections \cite{Bengtsson1999:PRB} were also included to get correct band offsets and internal electric fields.
To determine the interlayer distances, the atoms of BLG were allowed to relax only in their $z$ positions 
(vertical to the layers), and the atoms of CGT were allowed to move in all directions,
until all components of all forces 
were reduced below $10^{-3}$~[Ry/$a_0$], where $a_0$ is the Bohr radius.

\section{Electronic band structure of BLG on CGT.}
The electronic structure of bare BLG contains four parabolic bands near the Fermi energy \cite{Nilsson2008:PRB, McCann2013:RPP}. 
Two low energy bands close to the charge neutrality point are 
formed by ($p_z$) orbitals of non-dimer atoms B$_1$ and A$_2$, see Fig.~\ref{Fig:scheme}. 
In addition, there are two higher lying bands formed by the orbitals of dimer atoms A$_1$ and
B$_2$, see Fig.~\ref{Fig:scheme}. Since the dimer atoms are connected by direct
interlayer hopping, the bands are shifted roughly 400~meV from the Fermi level, such that 
they can be ignored for transport.

\begin{figure}[htb]
\centering
 \includegraphics[width=0.99\columnwidth]{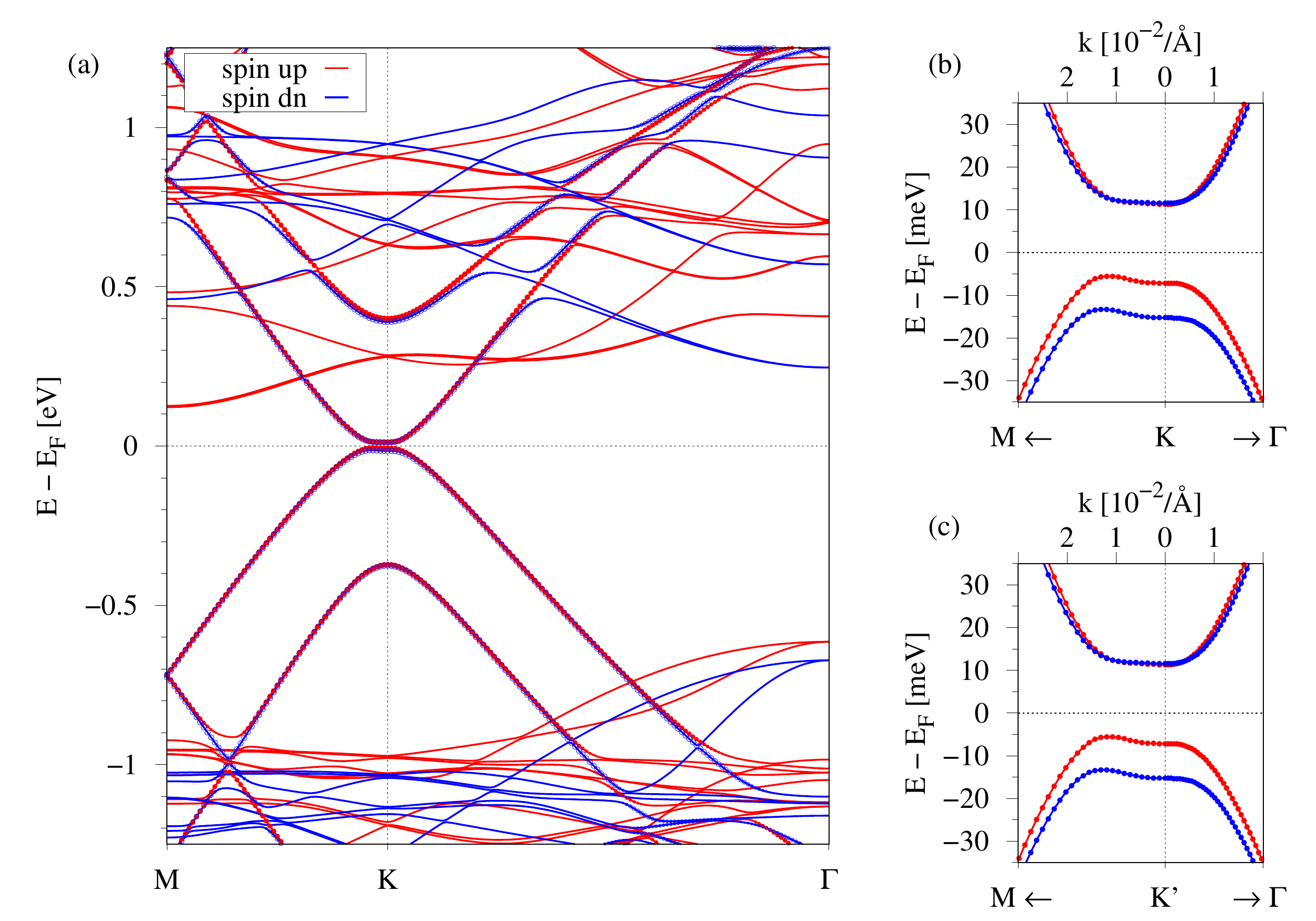}
 \caption{(Color online) (a) Calculated electronic band structure of bilayer graphene
 on monolayer Cr$_2$Ge$_2$Te$_6$. Bands of bilayer graphene are highlighted by blue and red circles. 
 (b) Zoom to the fine structure of the low energy bands at K point
 close to the Fermi level. (c) Same as (b), but for K' point. 
 Bands with spin up (down) are shown in red (blue).
 }\label{Fig:bands_BLG_CGT}
\end{figure}

In Fig.~\ref{Fig:bands_BLG_CGT}(a) we show the calculated electronic band structure
of BLG on monolayer CGT along high symmetry lines. 
The bands near the Fermi level resemble closely the 
bands from bare BLG~\cite{Nilsson2008:PRB, McCann2013:RPP}, 
even though the high energy conduction band of BLG is located within the conduction bands of CGT. 
Very important for transport is that the low energy bands of BLG are located
within the band gap of CGT. The offset between the conduction band of BLG at K and the
bottom of conduction band minimum of CGT is roughly $100$~meV.

However, there are two important differences compared to bare BLG. 
First, the heterostructure possesses an intrinsic 
dipole (vertical to the sheets) and thus the two graphene layers are at a different
potential energy resulting in a small indirect orbital gap of roughly $17$~meV, see Fig. \ref{Fig:bands_BLG_CGT}(b). 
The dipole of $1.505$~Debye points from CGT towards BLG and therefore 
B$_1$ (A$_2$) electrons have lower (higher) energy forming the valence (conduction) states. 
Second, the ferromagnetic CGT substrate interacts mainly with the lower graphene layer 
and, by proximity exchange, splits the low energy bands
originating from this graphene layer by $8$~meV, see Fig.~\ref{Fig:bands_BLG_CGT}(b). 
The upper graphene layer is far away and experiences almost no proximity effect, 
resulting in a comparatively small spin splitting of the corresponding bands. 
The calculated magnetic moment of the Cr-atoms is positive and roughly $3.2~\mu_{\rm{B}}$, but
the Te-atoms and consequently the C-atoms are polarized with a small negative magnetic moment. 
Thus the BLG low energy valence band is spin split, with spin down states being 
lower in energy, see Fig.~\ref{Fig:bands_BLG_CGT}(b).
We would also like to emphasize, that due to the ferromagnetic substrate we break time-reversal symmetry. 
Thus the band structure and also the splitting is the same for K and K' point, see Fig.~\ref{Fig:bands_BLG_CGT}(b,c). 
This is in contrast to proximity SOC in BLG on WSe$_2$ \cite{Gmitra2017:PRL}, 
where K and K' point are connected by time-reversal symmetry. 
\section{Proximity exchange valve}

\begin{figure}[htb]
\centering
 \includegraphics[width=0.9\columnwidth]{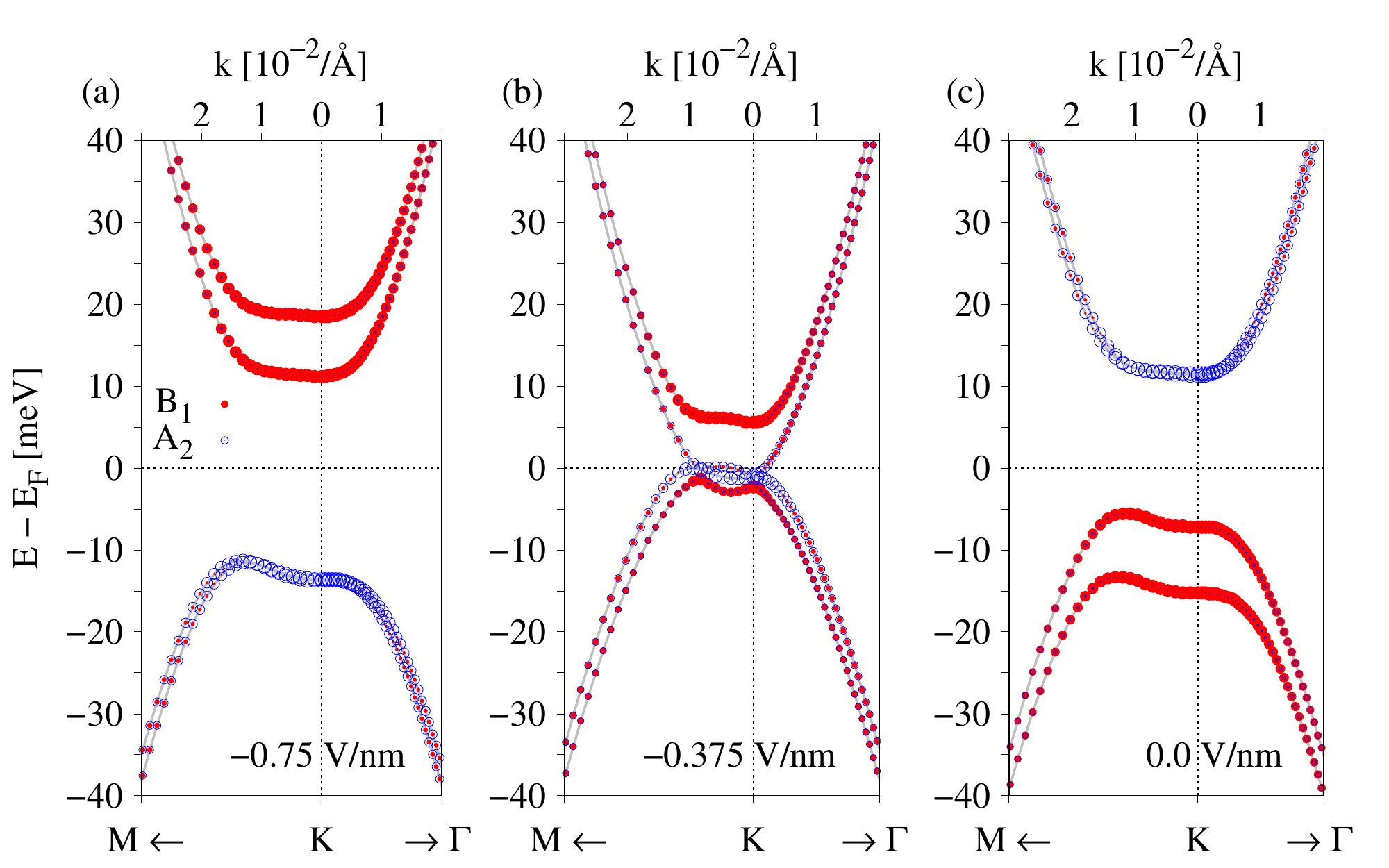}
 \caption{(Color online) Calculated sublattice resolved band structures 
 around K valley for vertical electric field of 
 (a)~{$-0.75$}~V/nm, (b)~{$-0.375$}~V/nm, (c)~{$0.0$}~V/nm.
 The circles radii correspond to the probability of the state 
 being localized on carbon atoms B$_1$ (red filled circles) and atoms A$_2$ (blue open circles).
 }\label{Fig:bands_chain}
\end{figure}

In order to design devices for spintronics it is desirable if one can electrically
control both spin and orbital properties. Such a control should
be highly efficient in our heterostructures. In Fig.~\ref{Fig:bands_chain} we show the low energy
band structures of BLG, in the presence of a vertical electric field. 
At zero field strength, see Fig.~\ref{Fig:bands_chain}(c),
the situation is as explained above; the hole band formed by $p_z$-orbitals of B$_1$ is strongly split 
by proximity exchange, while the electron band, formed by $p_z$-orbitals of A$_2$ is much 
less split. 
The reason is simply because the atoms B$_1$ are closer to the ferromagnetic substrate experiencing 
a stronger proximity effect than atoms A$_2$.
The ordering of the bands is determined by the built-in electric field, pointing from 
CGT towards BLG, and thus B$_1$ electrons are at lower energy.

Now, if we apply a positive external electric field, 
it adds to the internal field, opening the orbital gap further, keeping the band-dependent 
exchange splitting unchanged. But when the applied field is negative and strong enough to compensate
the built-in electric field, the orbital gap closes, see Fig.~\ref{Fig:bands_chain}(b). 
This compensation happens at roughly $-0.375$~V/nm and the states 
from atoms B$_1$ and A$_2$ are almost at the same potential energy, closing the gap. 

A further increase of the negative field leads to the 
reopening of the band gap, see Fig.~\ref{Fig:bands_chain}(a), but
with a switched character of the bands. 
Now, the conduction band is formed by atoms B$_1$, which still 
experience the stronger proximity effect, while the valence band is formed by atoms A$_2$,
which are far from the ferromagnetic substrate.
The switching of the band character happens, because the total (external and built-in)
field is pointing now from BLG towards CGT, opposite to the zero-field case.
Thus, atoms B$_1$ experience a higher potential than atoms A$_2$. 
We get a proximity exchange valve.

\begin{figure}[htb]
\centering
 \includegraphics[width=0.85\columnwidth]{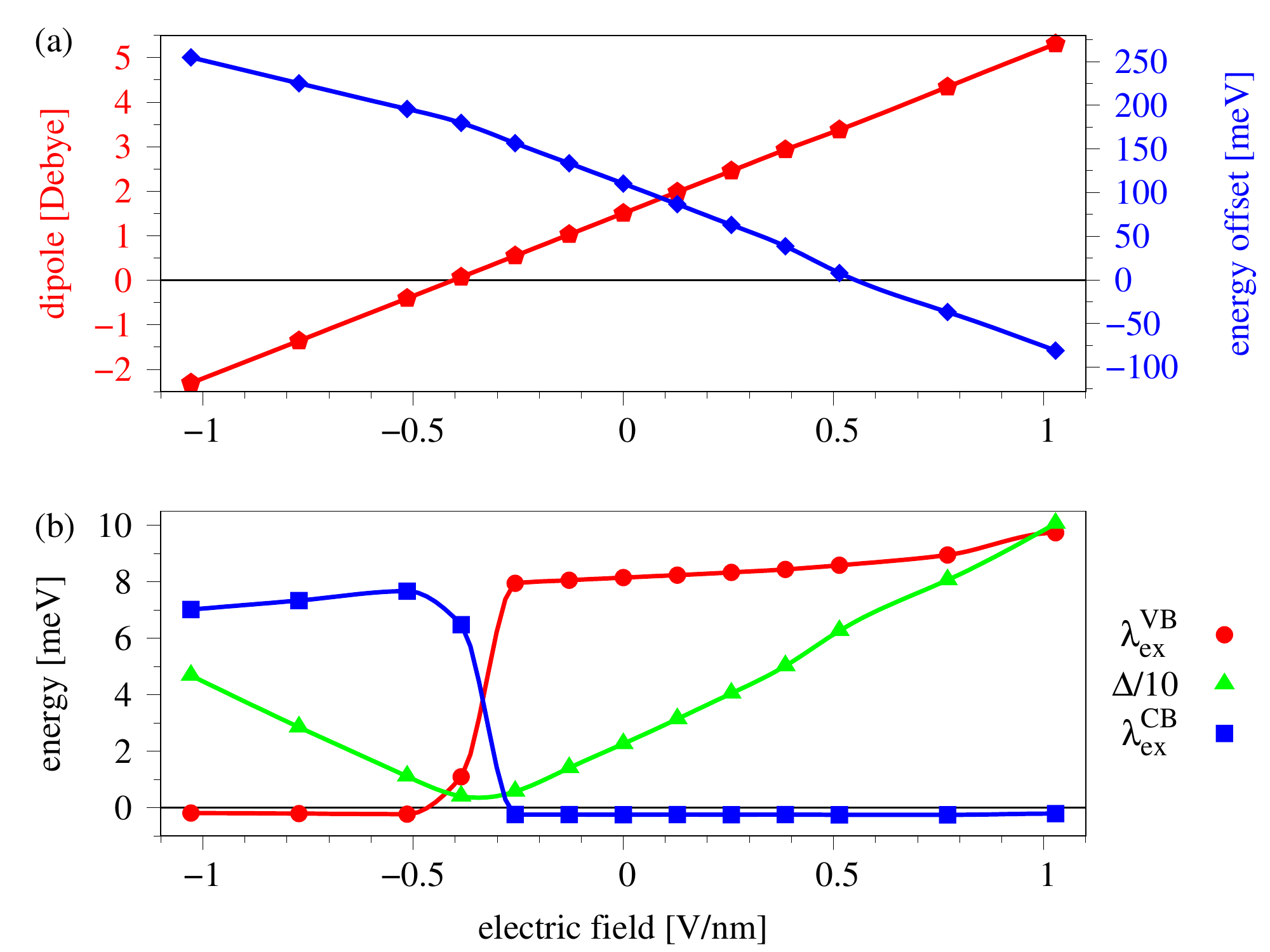}
 \caption{(Color online) Calculated electric field dependencies in bilayer graphene on monolayer Cr$_2$Ge$_2$Te$_6$ of
(a)~dipole induced in the heterostructure (red), and
energy offset (blue) of the bilayer graphene conduction band at K 
point and bottom of the conduction band minimum of the Cr$_2$Ge$_2$Te$_6$. 
(b)~Calculated band gap $\Delta$, valence band splitting $\lambda_{\textrm{ex}}^{\textrm{VB}}$ 
and conduction band splitting $\lambda_{\textrm{ex}}^{\textrm{CB}}$ at the K point as a 
function of the applied electric field. 
 }\label{Fig:dipole}
\end{figure}

Are the relevant BLG low-energy bands still within the orbital band gap of CGT?
In Fig.~\ref{Fig:dipole}(a), we see that 
the calculated dipole depends linearly on the applied electric field. The built-in dipole is roughly compensated by 
an external field of $-0.4$~V/nm; the amplitude of the intrinsic field is $0.406$~V/nm. 
Figure~\ref{Fig:dipole}(a) also shows that the energy offset between the BLG conduction band at K and 
the bottom of the conduction band minimum of CGT also changes with electric field. However, it is
important to note that the low energy states of BLG are located within the band gap of CGT for
electric fields that allow to observe the proximity exchange switching of the states.
For fields larger than $0.5$~V/nm, the conduction band of BLG at K is shifted above the bottom of the
conduction band minimum of CGT and BLG gets hole doped.

In Fig.~\ref{Fig:dipole}(b), we show the evolution of the
exchange splittings and the orbital gap as a function of the applied electric field.
For this we define the orbital gap $\Delta$,
as the average of the spin up ($\Delta_{\uparrow}$) and spin down ($\Delta_{\downarrow}$) orbital gaps,
$\Delta = (\Delta_{\uparrow} + \Delta_{\downarrow})/2$ at the K point. 
The exchange splitting is defined as the energy difference between spin up and spin down band,
$\lambda_{\textrm{ex}}= E_{\uparrow}- E_{\downarrow}$, at the K point for 
conduction band (CB) and valence band (VB) respectively. 
The orbital gap, which is a measure for the strength of the intrinsic dipole across the junction, 
is consistent with our dipole values. In Fig.~\ref{Fig:dipole}(b), we see that the gap $\Delta$ has its minimum value, 
when the applied field compensates the intrinsic dipole, at around $-0.4$~V/nm. 
If we look at the two splitting parameters $\lambda_{\textrm{ex}}^{\textrm{VB}}$ 
and $\lambda_{\textrm{ex}}^{\textrm{CB}}$, we can see that there is a very nice switching behavior 
around $-0.4$~V/nm, where the gap $\Delta$ has its minimum.
The conduction band splitting $\lambda_{\textrm{ex}}^{\textrm{CB}}$ is around $7.5$~meV 
for fields smaller than $-0.4$~V/nm and then drops to a small negative value when applying a larger field strength. 
The opposite behavior is observed for the valence band splitting $\lambda_{\textrm{ex}}^{\textrm{VB}}$, 
which has a large magnitude for fields larger than $-0.4$~V/nm, see Fig.~\ref{Fig:dipole}(b).
We find that the switching happens within a few hundred millivolts per nanometer, 
and most important at experimentally accessible field strengths.

\section{BLG on \texorpdfstring{C\lowercase{r}$_2$S\lowercase{i}$_2$T\lowercase{e}$_6$}{} and \texorpdfstring{C\lowercase{r}$_2$S\lowercase{n}$_2$T\lowercase{e}$_6$}{}{}}
\begin{figure}[htb]
\centering
 \includegraphics[width=0.99\columnwidth]{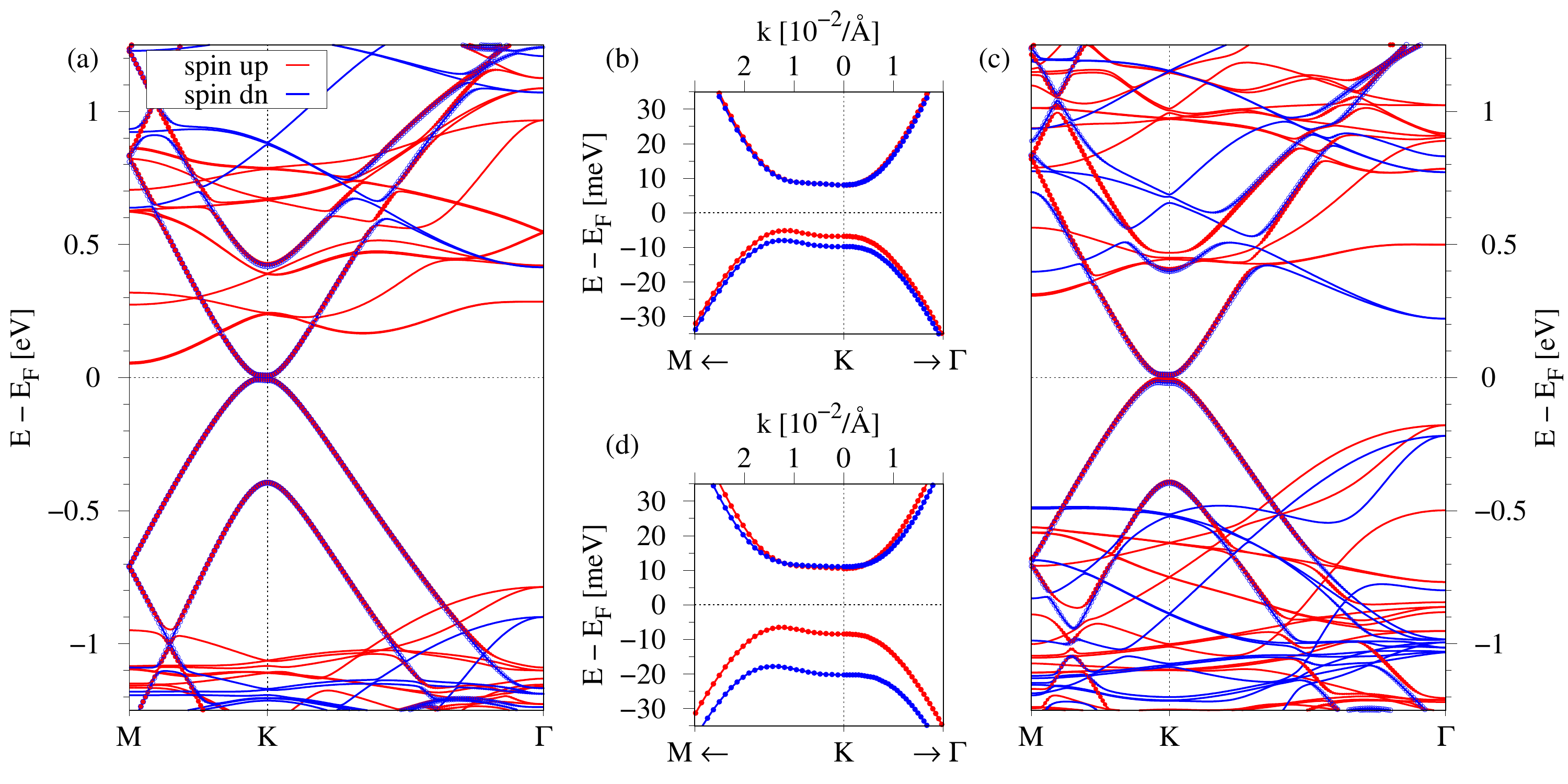}
 \caption{(Color online) (a) Calculated electronic band structure of bilayer graphene
 on monolayer Cr$_2$Si$_2$Te$_6$. Bands of bilayer graphene are highlighted by blue and red circles.
 (b)~Zoom to the fine structure of the low energy bands close to the Fermi level. 
 Bands with spin up (down) are shown in red (blue). (c) and (d) are the
 same as (a) and (b), but for bilayer graphene on Cr$_2$Sn$_2$Te$_6$.
  }\label{Fig:bands_BLG_CrSiSnTe}
\end{figure}

We predict similar scenarios for X=Si and Sn. In Fig. \ref{Fig:bands_BLG_CrSiSnTe}  
we show the band structure of BLG on Cr$_2$Si$_2$Te$_6$ and Cr$_2$Sn$_2$Te$_6$. 
Very similar to the bands of BLG on CGT, we see proximity spin splitting in the low energy valence band, 
much larger than in the conduction band. 
Here we would like to mention that the compounds with Si and Ge have been identified in 
experiments~\cite{Carteaux1991:JMM, Carteaux1995:JP}, 
whereas the compound with Sn atoms has been only predicted theoretically~\cite{Zhuang2015:PRB} so far.

In the Si (Sn) case, proximity induced exchange is smaller (larger), as the valence band spin splitting is
about $3$~meV ($12$~meV) at the K point and the indirect gap is roughly $13$~meV ($17$~meV).
The energy offset between the conduction band of BLG at K and the
bottom of conduction band minimum of Cr$_2$Si$_2$Te$_6$ (Cr$_2$Sn$_2$Te$_6$) is roughly $44$~meV ($211$~meV).
The intrinsic dipole is weaker (stronger), about $1.336$~($2.013$)~Debye, which in the end
leads to slightly different electric fields to observe the switching of the band character, 
as shown for BLG on CGT. 
The average distance between the lower graphene layer and Cr$_2$Si$_2$Te$_6$ (Cr$_2$Sn$_2$Te$_6$)
was relaxed to $3.563$~($3.531$)~\AA.
From the minimal difference in distance, we conclude that the spin splitting strongly depends on the 
material itself, whereas in Ref. \cite{Zhang2015:PRB}, it was already shown that proximity induced
exchange splitting strongly depends on the distance between graphene and the CGT substrate.

For model transport calculations it is useful to have realistic parameters in order to estimate
the conductance through BLG in this heterostructures. 
In Tab.~\ref{tab:parameters} we summarize the most relevant parameters 
for bilayer graphene on all three considered substrates, for zero applied external field.
The exchange splitting of the valence band is always roughly $30$ times larger than the splitting of
the conduction band, independent of the material. Note that $\lambda_{\textrm{ex}}^{\textrm{CB}}$ has the opposite sign of 
$\lambda_{\textrm{ex}}^{\textrm{VB}}$, see Tab.~\ref{tab:parameters}, 
meaning that the order of the spin bands for conduction band is different from the valence band, at the K point.

\begin{table*}[htb]
\caption{\label{tab:parameters}
Summary of parameters for bilayer graphene on monolayer Cr$_2$X$_2$Te$_6$ for zero field and X = Si, Ge, and Sn. 
The parameter $\Delta$ describes the average orbital gap in the spectrum at K, $\lambda_{\textrm{ex}}$ are the 
exchange splittings for valence band (VB) and conduction band (CB) at K. The strength of the calculated 
intrinsic dipole across the heterostructure is given in debye,
and $d$ is the average distance between the lower graphene layer and the substrate.
The lattice constant $a_0$ is the one from the ferromagnetic substrate. 
In order to match the $5\times 5$ graphene unit cell
on the FMI we had to stretch its lattice constant to $7.1014$~\AA. The strain gives, by how much
we stretch the substrate. Finally $T_c$ is the Curie temperature, either from experiment (exp.) on bulk samples or from 
Monte-Carlo simulations (theo.) on monolayers.
}
\begin{indented}
\setlength{\tabcolsep}{3.5mm}
\renewcommand{\arraystretch}{1.3}
\item[]\begin{tabular}{llll}
\br
& Cr$_2$Si$_2$Te$_6$ & Cr$_2$Ge$_2$Te$_6$ & Cr$_2$Sn$_2$Te$_6$\\
\mr
\textrm{$\Delta$} [meV]& 16.15 & 22.55 & 25.15 \\
\textrm{$\lambda_{\textrm{ex}}^{\textrm{VB}}$} [meV]& 3.0 & 8.2 & 11.8 \\ 
\textrm{$\lambda_{\textrm{ex}}^{\textrm{CB}}$} [meV]&-0.1 & -0.3 & -0.5 \\
\textrm{dipole} [debye]& 1.336 & 1.505 & 2.013 \\
\textrm{$d$} [\AA]& 3.563 & 3.516 & 3.531 \\
\textrm{$a_0$} [\AA] & 6.758~\cite{Carteaux1991:JMM} & 6.828~\cite{Carteaux1995:JP} & 7.010~\cite{Zhuang2015:PRB}\\
\textrm{strain} [\%] & 5.1 & 4.1 & 1.3\\
\textrm{$T_c$} [K] & 33 (exp.~\cite{Carteaux1991:JMM}), & 61(exp.~\cite{Carteaux1995:JP}), & 170 (theo.~\cite{Zhuang2015:PRB})\\
& 90 (theo.~\cite{Zhuang2015:PRB}) & 130 (theo.~\cite{Zhuang2015:PRB}) & \\
\br
\end{tabular}
\end{indented}
\end{table*}

\section{Bilayer graphene device}
The electrical switching of the exchange splitting is a nice platform to realize 
novel spintronics devices, particularly since high spin-polarization can be achieved. Indeed,
without any applied field, the chemical potential can be tuned close to the maximum of the valence
band where states are 100\% spin polarized. The same would hold for the conduction band
in an applied negative field. Based on theoretical model calculations 
--- on magnetoresistance in BLG \cite{Semenov2008:PRB} and 
spin transport in graphene \cite{Haugen2008:PRB,Semenov2010:JAP}, both being subject to proximity exchange --- 
model transport calculations in FMI/BLG were performed \cite{Michetti2010:NL, Michetti2011:PRB}. 
It was found that the proximity induced spin splitting 
allows BLG to act as a spin filter and spin rotator, being electrically controllable. 

\begin{figure}[htb]
\centering
 \includegraphics[width=0.8\columnwidth]{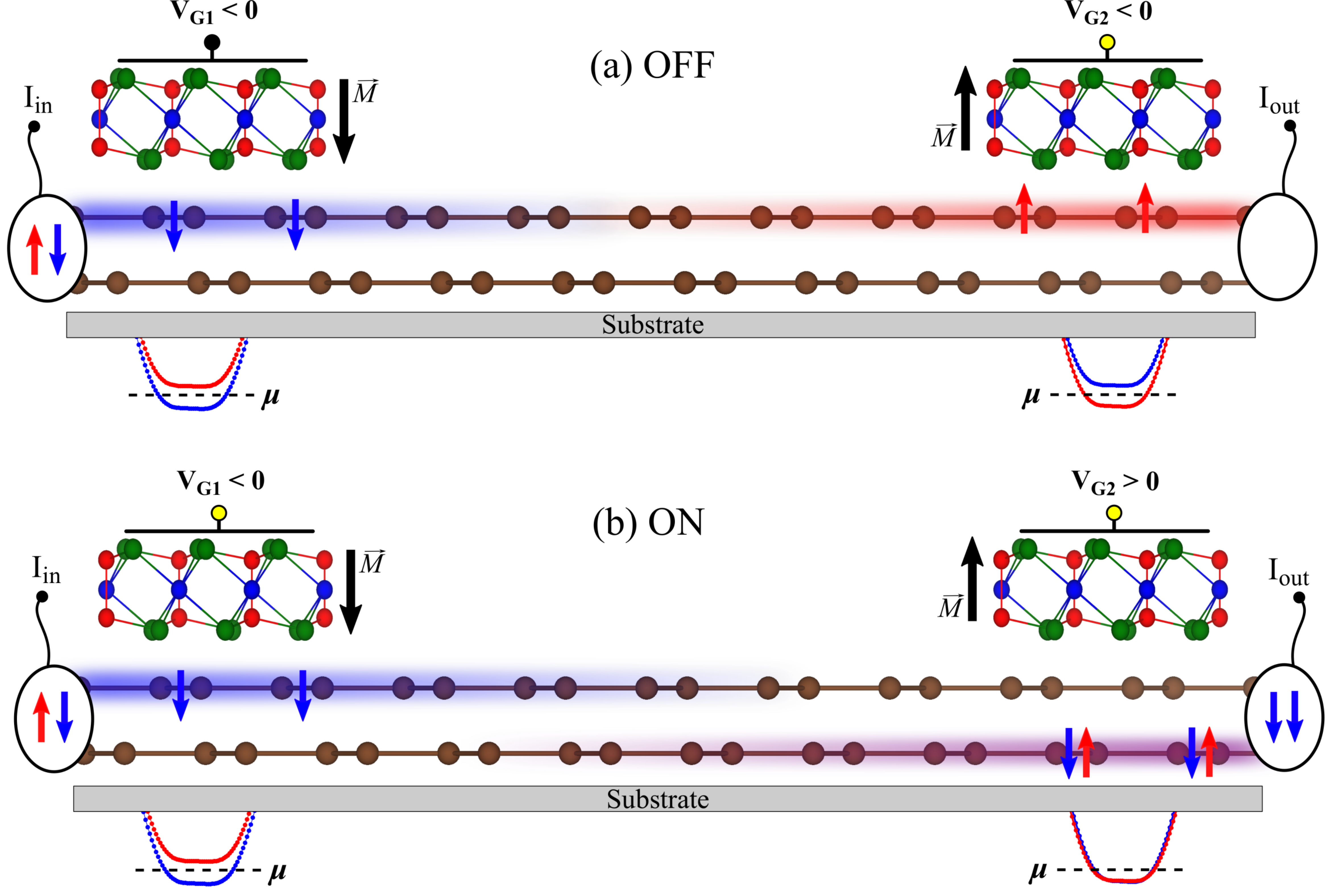}
 \caption{(Color online) Schematics of a spin filter device, consisting of BLG on a substrate with two
 regions of a FMI (such as CGT) on BLG, where the two
 FMIs have opposite magnetization aligned along $-z$ and $z$. Below each heterostructure region
 the low energy conduction bands are depicted. Spins are shown as red (spin up) and blue (spin down) arrows.
 Depending on the gate voltage $\rm{V}_{\rm{G2}}$ we can switch into (a) OFF state and (b) ON state. 
 }\label{Fig:spin_filter}
\end{figure}

In Fig. \ref{Fig:spin_filter} we show the structure of a spin filter, where BLG is deposited on an insulating substrate.
On top of BLG we have two regions where CGT is placed, but in principle also other FMIs (YIG, EuO, CrI$_3$) are possible candidates
to realize the following devices. On top of the regions with the FMIs, gate 
voltages $\rm{V}_{\rm{G1}}$ and $\rm{V}_{\rm{G2}}$ can be applied.
The magnetization directions $\bm{M}$ of the two FMIs are opposite. 

Suppose the gate voltage $\rm{V}_{\rm{G1}}$ is negative, 
then the low energy conduction bands of BLG below that first FMI are formed by the upper graphene layer
and are strongly spin split, see Fig. \ref{Fig:spin_filter}(a). 
A spin unpolarized current is injected from the left and enters the first heterostructure. 
If the chemical potential $\mu$ is between the two spin split electron bands, one spin component is
filtered. In the central part we have just bare BLG where, due to small intrinsic SOC, the spin
keeps its direction aligned until it enters the second heterostructure. 
Depending on the gate voltage $\rm{V}_{\rm{G2}}$, we can have a ON or an OFF state. 
Since the magnetization of the second FMI is opposite to the first one the electrons
cannot travel on in the upper layer if $\rm{V}_{\rm{G2}}$ is also negative, 
because only the opposite spin channel is open, see Fig. \ref{Fig:spin_filter}(a). 
However, if $\rm{V}_{\rm{G2}}$ is positive, the conduction bands are formed by the lower
graphene layer, having almost no proximity exchange, and the electrons have both 
spin channels open, see Fig. \ref{Fig:spin_filter}(b). 

\begin{figure}[htb]
\centering
 \includegraphics[width=0.8\columnwidth]{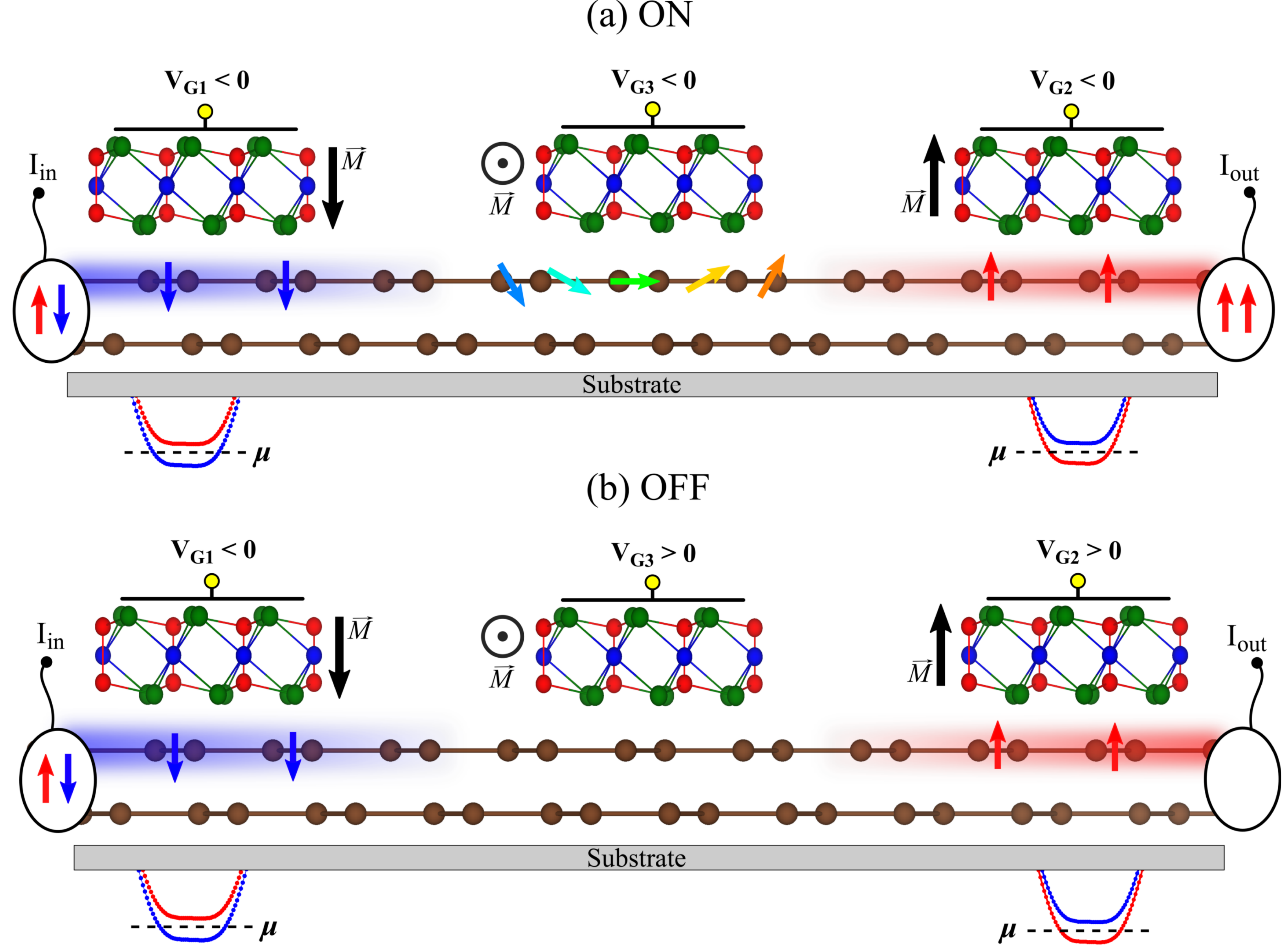}
 \caption{(Color online) Schematics of the spin rotator device. 
 The magnetization directions of the FMIs from left to right are $-z,~y,~z$ and spin 
rotation happens below the central FMI. Depending on the gate voltage $\rm{V}_{\rm{G3}}$ we can
switch (a) ON or (b) OFF the spin rotation of the conduction electrons.
 }\label{Fig:spin_rotator}
\end{figure}
A different approach is based on the spin rotation, see Fig. \ref{Fig:spin_rotator}. 
Similar to the spin filter, we polarize first a spin unpolarized current injected from the left, which is then 
analyzed in the right heterostructure. 
The difference is that there is an additional FMI able to control the carrier spins
via gate voltage $\rm{V}_{\rm{G3}}$. 
The magnetization directions of the FMIs from left to right are $-z,~y,~z$ and spin 
rotation happens below the central FMI. 
Depending on the gate voltage $\rm{V}_{\rm{G3}}$, carriers will flip their spin in the central region
and the channel on the right is open, see Fig. \ref{Fig:spin_rotator}(a), 
or closed, see Fig. \ref{Fig:spin_rotator}(b).
The origin of spin rotation and the efficiency of this device has been 
discussed in Refs. \cite{Michetti2010:NL, Michetti2011:PRB}, and is based on wave function localization
and short ranged proximity exchange. 

Above are but a few examples of spin devices based on the field-effect exchange coupling concepts. BLG 
on ferromagnetic insulating substrates could be employed to build bipolar spintronics elements, such as spin 
diodes and spin transistors~\cite{Zutic2006:IBM} or spin logic devices~\cite{Behin2010:NN}, 
which require exchange coupling (but non necessarily magnetic moments) to 
make their electronic bands spin dependent.

\section{Summary and Conclusion}
In conclusion, we have studied from first principles
the electronic structure of bilayer graphene on Cr$_2$Ge$_2$Te$_6$. 
We have shown that we can efficiently, and most important fully electrically, switch the
exchange interaction by one order of magnitude, of electrons and holes. At low enough energy
the electronic states can be 100\% spin polarized, which can lead to interesting new
device concepts. We also expect that our DFT-derived parameters will be useful in model
transport calculations involving exchange proximitized BLG.


\section*{Acknowledgments}
This work was supported by DFG SPP~1666, SFB~689, SFB~1277 (A09 and B07), and by the European 
Unions Horizon 2020 research and innovation
programme under Grant agreement No. 696656.
The authors gratefully acknowledge the Gauss Center for Supercomputing e.V. 
for providing computational resources on the GCS Supercomputer SuperMUC at Leibniz Supercomputing Center.\\

\section*{References}
\bibliographystyle{iopart-num}
\bibliography{paper}

\end{document}